# An adaptive algorithm for embedding information into compressed JPEG images using the QIM method

Anna Melman
*Department of Information Systems Security*
Tomsk State University of Control Systems and Radioelectronics
Tomsk, Russia
anna.s.kokurina@tusur.ru

Pavel Petrov
*Department of Complex Information Security of Computer Systems*
Tomsk State University of Control Systems and Radioelectronics
Tomsk, Russia
725_ppo@fb.tusur.ru

Alexander Shelupanov
*Department of Complex Information Security of Computer Systems*
Tomsk State University of Control Systems and Radioelectronics
Tomsk, Russia
saa@fb.tusur.ru

*Abstract*— The widespread use of JPEG images makes them good covers for secret messages storing and transmitting. This paper proposes a new algorithm for embedding information in JPEG images based on the steganographic QIM method. The main problem of such embedding is the vulnerability to statistical steganalysis. To solve this problem, it is proposed to use a variable quantization step, which is adaptively selected for each block of the JPEG cover image. Experimental results show that the proposed approach successfully increases the security of embedding.

## I. INTRODUCTION

Digital steganography is one of modern areas of information security. It solves the same urgent problem as cryptography - ensuring secure transmission of information. Unlike cryptography, it does not change the information itself, making it unreadable, but ensures the creation of a covert data transmission channel by embedding messages in digital objects. In recent years, various steganography algorithms have been actively created and investigated [1, 2].

A significant part of steganographic algorithms deal with digital images, since they are widespread and allow transferring secret information without undue suspicion. This research also focuses on hiding information in digital images. JPEG images are considered as cover images for secret messages, since a large number of images are stored and transmitted on the network in the JPEG format.

Steganographic embedding should be invisible, including for steganalysis methods. Therefore, the purpose of this work is to develop a data hiding algorithm that ensures the invisibility of embedding. The paper proposes an adaptive algorithm for embedding information into compressed JPEG images, based on the well-known steganographic method of quantization index modulation (QIM) and allowing to reduce the vulnerability of embedding to steganalysis, in particular, histogram analysis.

## II. JPEG COMPRESSION METHOD

JPEG is one of the most popular and widespread lossy compression methods for images. The file size is reduced by removing some of the redundant information from the image. The user-defined compression ratio determines the visual quality of the image after compression.

The main stages of JPEG compression are converting the color space of the image to YCbCr, "decimating" the Cb and Cr channels, discrete cosine transform (DCT), quantization and encoding. Stages of DCT and quantization are of the greatest interest from the point of view of steganographic embedding of information.

DCT is one of the most common frequency transformations. The result of applying DCT to the matrix of intensities of image pixel values is a frequency coefficients matrix of the same size. In the JPEG standard, this transformation is applied by 8x8 matrices. The coefficient in the upper left corner is called the DC coefficient, the rest of the coefficients are called AC coefficients. The most significant information is contained in the low frequency (closer to the top left corner) coefficients. Mid and high frequency (closer to the lower right corner) coefficients are less important for the subsequent image reconstruction.

The compression occurs at the quantization stage, when each DCT coefficient is divided by some value determined by the quantization matrix and rounded. As a result an irreversible loss of information occurs.

Some algorithms for embedding information in JPEG images embed message bits in DCT coefficients before quantization or combine embedding with a quantization procedure, for example, algorithm [3]. However, in most cases, the quantized AC coefficients are changed. In this work, this particular approach is implemented.

### III. QIM EMBEDDING METHOD

The QIM method [4] is one of the popular methods for embedding information into digital images. Its main idea is to change the image data item (pixel value or frequency coefficient) depending on the bit value of the secret message. The image data item value to be changed is divided by a special coefficient and then rounded off. This coefficient is called the quantization step $q$. In this case, the message bit embedding formula has the following form:

$$c' = q \cdot \left\lfloor \frac{c}{q} \right\rfloor + \frac{q}{2} \cdot b_i, \qquad (1)$$

where $c$ is a DCT coefficient value before embedding, $c'$ is a DCT coefficient value after embedding, $q$ is the quantization step, $b_i$ is a message bit, $\lfloor ... \rfloor$ is an operation of obtaining integer part from division.

The extraction is performed using the formula

$$b'_i = \arg\min_{p \in [0,1]} \left| c'' - c'_p \right|, \qquad (2)$$

where $c''$ is a DCT coefficient containing the message bit, $c''_0 = q \cdot \left\lfloor \frac{c''}{q} \right\rfloor$, $c''_1 = q \cdot \left\lfloor \frac{c''}{q} \right\rfloor + \frac{q}{2}$.

The efficiency of embedding is mainly determined by the size of the quantization step. The larger the $q$ value, the greater the resistance of the embedding to various distortions, but at the same time the greater the vulnerability to detecting the presence of the embedding using steganalysis, and sometimes even with the naked eye.

### IV. ADAPTIVE SELECTION OF THE QUANTIZATION STEP

Information embedding in JPEG images using the QIM method lead to significant distortions of the natural model of a digital image in the frequency domain. In the context of this work, the histogram of quantized AC coefficients of the image is used as such a model.

To illustrate this statement, let us turn to Fig. 1, which shows examples of histograms before (Fig.1a) and after (Fig.1b) embedding information into a JPEG image. Obviously, these histograms differ significantly. The histogram shown in Fig. 1b contains characteristic "dips", because due to the application of formula (1), AC coefficients frequency values change.

This is a usual problem for the classical QIM method. An effective way to solve this problem is to use a variable quantization step for each block of the image. For example, in [3], such an approach allows for the statistical invisibility of embedding. However, the disadvantage is the need to use some additional information when embedding. In particular, in this case we are talking about auxiliary sequences for generating quantization steps. Sending such information apart from the stego image poses a threat to the attachment being detected. Additional information can draw the attention of an attacker and inform him of the presence of a hidden data transmission channel. If the additional information is unique for each container-attachment pair, then for each communication session it is necessary to solve the problem of its protection, for example, by encryption. In this case, the use of steganography becomes impractical, since it is more convenient to apply encryption to the secret message itself.

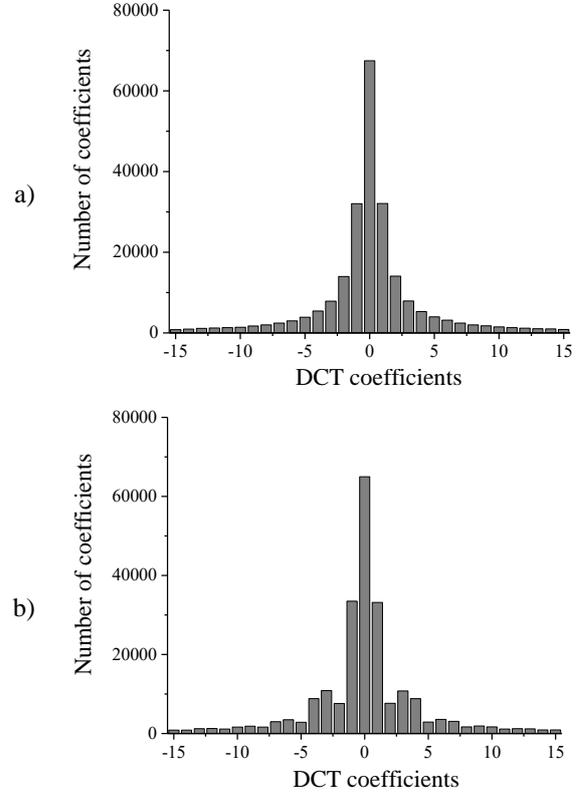

Fig. 1. Histograms of AC coefficients: a) before embedding, b) after embedding.

Note that the transmission of key information, for example, algorithm parameters, does not pose the same serious problem, since it is chosen once for a group of users.

In the article [5] of one of the authors of this study, a solution was proposed that was free of this drawback. It was proposed to use a part of the DCT coefficients of the block that were not used to hide the message (for uncompressed images) to select the quantization step. The quantization step was defined as the smallest value of the least frequently encountered coefficients in the non-embedding area.

The non-embedding area is a range of coefficients that do not change when information is embedded. The embedding area is the range of coefficients that can be changed. Embedding information in the mid- and high-frequency DCT coefficients leads to less distortion of the stego image, therefore they are the area of embedding. The rest of the AC-coefficients constitute the non-embedding area. An example of an embedding and non-embedding area is shown in Fig. 2.

In this paper, it is also proposed to choose the smallest value of the quantized AC coefficient of the non-embedding area from the rarest ones to select the quantization step for each block of a JPEG image. This helps to redistribute the "dips" in the histogram and improve the security of the embedding. In this case, the choice of the quantization step in the non-embedding area has two advantages. First, it makes the embedding adaptive; taking into account the characteristics of a particular cover image. Secondly, it makes it possible to refuse the transfer of additional information, unique for each case of embedding, which serves as an unmasking feature of the embedding.

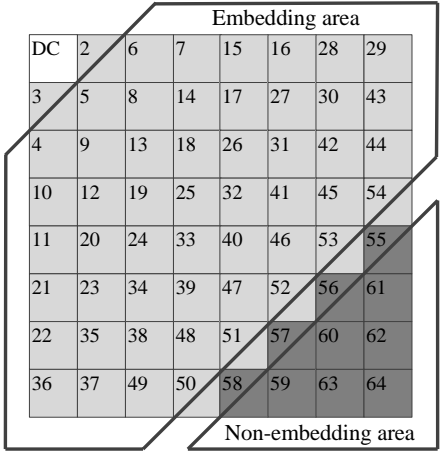

Fig. 2. Embedding area and non-embedding area in the block of DCT coefficients.

## V. INFORMATION EMBEDDING ALGORITHM

Thus, let us formulate the main stages of the algorithm for embedding information into JPEG images that implements the described approach.

Algorithm input: JPEG cover image, secret message.

Algorithm output: stego image.

Step 1: Get the quantized DCT coefficients from the JPEG image.

Step 2: For each block of the cover image, do the following:

Step 2.1: Determine the value of a quantization step $q$ by the non-embedding area.

Step 2.2: Separate a message fragment equal to the number of coefficients in the non-embedding area.

Step 2.3: Embed a message fragment into a block using formula (1) with the quantization step $q$.

Step 3: Encode the coefficients and generate the stego image.

When extracting information, the message bits are extracted according to formula (2) from the embedding area after preliminary calculation of the quantization step $q$ over the non-embedding area. Since the non-embedding area does not change when the secret message is hidden, the $q$ value is the same, and the secret message is extracted without any error.

## VI. EXPERIMENTAL RESULTS

To assess the effectiveness of the developed algorithm, computational experiments were performed. For the experiments, we used 10 images from the USC-SIPI database [6] with a resolution of 512 × 512. Each image has been compressed in JPEG format with quality factor 95.

The peak signal-to-noise ratio (PSNR) metric was used to numerically evaluate the visual quality of stego images. Fig. 3 shows a graph of the dependence of the PSNR value on the embedding capacity, averaged over the entire sample of images. Steganographic embedding is considered invisible to the human eye if the PSNR value is 30-35 dB. The proposed algorithm provides the required quality level. This achieves an acceptable level of maximum capacity of about 50,000 bits.

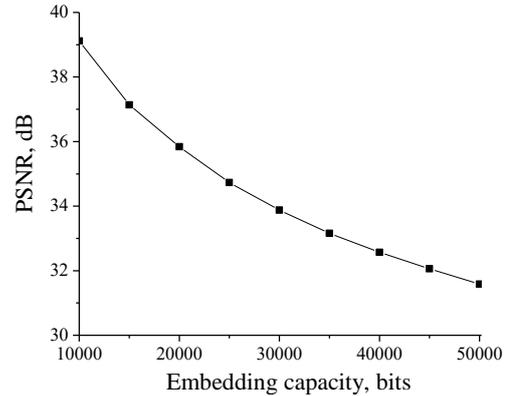

Fig. 3. Dependence of the PSNR value on the embedding capacity.

Figure: 4 shows a histogram of DCT coefficients of the stego image after embedding according to the described algorithm. Comparing Fig. 1b and Fig. 4, it is obvious that the adaptive choice of the quantization step allowed us to avoid significant distortions of the histogram.

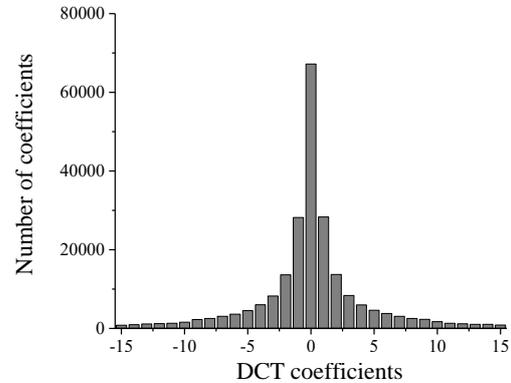

Fig. 4. Histogram of the stego image AC coefficients with a variable quantization step.

## VII. CONCLUSION

In this work, an algorithm for steganographic embedding of information into compressed JPEG images based on the QIM method was proposed. A distinctive feature of the proposed algorithm is the adaptive selection of the quantization step depending on each specific cover image. Experimental results show that the proposed algorithm is distinguished by increased resistance to the analysis of histograms. At the same time, the algorithm provides high visual quality of stego images.

ACKNOWLEDGMENT

This research was funded by the Ministry of Science and Higher Education of Russia, Government Order for 2020–2022, project no. FEWM-2020-0037 (TUSUR).